\documentclass[conference]{IEEEtran}
\usepackage[utf8]{inputenc}
\IEEEoverridecommandlockouts
\usepackage{url}
\usepackage{subfig}
\usepackage{amssymb}

\usepackage{url}
\usepackage{cite}
\usepackage{amsmath,amssymb,amsfonts}
\usepackage{algorithmic}
\usepackage{graphicx, url}
\usepackage{textcomp}
\usepackage{xcolor}
\usepackage{booktabs}
\usepackage{indentfirst}
\usepackage{fnpos}
\usepackage[T1]{fontenc}
\usepackage{float}
\usepackage{minted}
\usepackage{enumitem}
\usepackage{comment}
\usepackage{adjustbox}
\usepackage{caption}
\usepackage{booktabs}
\usepackage{subcaption}
\usepackage{multirow}
\usepackage{subfig}
\renewcommand{\figurename}{Figure}
\everymath{\displaystyle}
\usepackage{amsmath, amsthm, amssymb, amsfonts, dsfont}

\begin{document}

% % https://globecom2025.ieee-globecom.org/sites/globecom2025.ieee-globecom.org/files/GC25-CFP-SAC_AC.pdf
% The page length limit for all initial submissions for review is SIX (6) printed pages (10-point font) and must be written in English. Initial submissions longer than SIX (6) pages will be rejected without review.

\title{
% TODO: 8

Benchmarking Distributed Architectures for UTM: A Comparative Analysis of InterUSS and HLF
%Urban Air Mobility: when drones face hard times in throughput performance evaluation
%\thanks{The authors are supported in part by the grant \#2020/09850-0. São Paulo Research Foundation (FAPESP).}
}

%\title{Blockchain vs Federated Architectures for Drone Traffic Management: Performance Benchmark of Hyperledger Fabric and InterUSS}
%\title{When Drones Meet Blockchain: Performance Benchmark of Advanced Air Mobility Architectures}
%\title{Performance Benchmark of Decentralized Backend for Advanced Air Mobility}
%\title{Performance Benchmark of Permissioned Blockchain for Advanced Air Mobility}
%\title{Performance Benchmark of Permissioned Blockchain for Advanced Air Mobility: an UTM Analisys}
%\title{Permissioned Blockchain for Advanced Air Mobility: A Performance Benchmark for UTM}
%\title{Permissioned Blockchain for Advanced Air Mobility: A Performance Analisys for UTM}
\title{Permissioned Blockchain in Advanced Air Mobility: A Performance Analysis for UTM}

\author{
\IEEEauthorblockN{Rodrigo Nunes\IEEEauthorrefmark{1}, André Melo\IEEEauthorrefmark{1}, Rafael Albarello\IEEEauthorrefmark{1}, Reinaldo Gomes\IEEEauthorrefmark{2}, Cesar Marcondes\IEEEauthorrefmark{1}, Lourenço Pereira Jr\IEEEauthorrefmark{1}}
\IEEEauthorblockA{\IEEEauthorrefmark{1}Computer Science Division, Aeronautics Institute of Technology, Brazil}
\IEEEauthorblockA{\IEEEauthorrefmark{2}Brazilian National Research and Educational Network, RNP, Brazil\\
\{andre.melo,albarello,rodrigosergio\}@ita.br, reinaldo.gomes@rnp.br, \{cmarcondes,ljr\}@ita.br}
}

% \author{

% \IEEEauthorblockN{André Luiz Elias Melo}
% \IEEEauthorblockA{\textit{Division of Computer Science} \\
% \textit{Aeronautics Institute of Technology}\\
% Sao Jose dos Campos, SP, Brazil \\
% andre.melo@ga.ita.br}
% \and
% \IEEEauthorblockN{Rafael Hickmann Albarello}
% \IEEEauthorblockA{\textit{Division of Computer Science} \\
% \textit{Aeronautics Institute of Technology}\\
% Sao Jose dos Campos, SP, Brazil \\
% albarello@ita.br}
% \and
% \IEEEauthorblockN{Rodrigo Sergio dos Santos Nunes}
% \IEEEauthorblockA{\textit{Division of Computer Science} \\
% \textit{Aeronautics Institute of Technology}\\
% Sao Jose dos Campos, SP, Brazil \\
% rodrigosergio@ita.br}
% \and
% \IEEEauthorblockN{Reinaldo Cezar de Morais Gomes}
% \IEEEauthorblockA{
% \textit{Rede Nacional de Pesquisas}\\
% Rio de Janeiro, RJ -- Brazil\\
% reinaldo.gomes@rnp.br}
% \and
% \IEEEauthorblockN{Cesar Marcondes }
% \IEEEauthorblockA{\textit{Division of Computer Science} \\
% \textit{Aeronautics Institute of Technology}\\
% Sao Jose dos Campos, SP, Brazil \\
% cmarcondes@ita.br}
% \and
% \IEEEauthorblockN{Lourenco Alves Pereira Jr.}
% \IEEEauthorblockA{\textit{Division of Computer Science} \\
% \textit{Aeronautics Institute of Technology}\\
% Sao Jose dos Campos, SP, Brazil \\
% ljr@ita.br}
% }

\maketitle
%  precisamos de 6 páginas
\thispagestyle{plain} \pagestyle{plain} % Comando para enumerar paginas

\begin{abstract}
% TODO: 7

The integration of Uncrewed Aerial Vehicles (UAVs) into low-altitude airspace has led authorities to adopt distributed Uncrewed Traffic Management (UTM) architectures that ensure interoperability and safety. Blockchain has been proposed as an enabler for trustworthy coordination among UTM stakeholders. Yet, its real-time performance under aeronautical constraints remains insufficiently characterized. This paper presents a quantitative benchmark comparing two regulation-compliant distributed architectures: the federated InterUSS platform maintained by the Linux Foundation and a permissioned blockchain based on Hyperledger Fabric. Both systems were evaluated through Operational Intent Reference (OIR) registration workloads generated via Hyperledger Caliper, measuring throughput, latency, and transaction loss under loads up to 50 transactions per second. Results show that InterUSS sustained sub-second latency and stable performance up to 30 TPS. At the same time, Fabric exhibited exponential degradation with median latency exceeding 3\,s and tail latencies above 15\,s beyond that point. These findings demonstrate that blockchain-based architectures must be redesigned to meet aeronautical timing and scalability requirements, suggesting that hybrid models combining distributed ledgers for auditability with federated frameworks for real-time coordination are more suitable for future UTM deployments.

% The rapid adoption of Uncrewed Aerial Vehicles (UAVs) has driven aviation authorities to propose distributed Uncrewed Traffic Management (UTM) architectures. Several studies have advocated blockchain as a promising technology to meet these requirements. However, since UTM is a safety-critical and highly regulated domain, compliance with standards and regulatory frameworks is as crucial as performance and security. This work benchmarks two distributed architectures aligned with current regulatory frameworks: the Linux Foundation’s InterUSS platform and a Hyperledger Fabric-based private ledger. Our findings reveal that blockchain-based systems require architectures specifically designed for aeronautical performance constraints. 
%Up to 30 TPS, InterUSS sustained near-constant median latency (80-100 ms), while Hyperledger Fabric exhibited exponential growth (580-790 ms). Beyond this point, Fabric’s latency exceeded 3 s at 50 TPS and p90 > 15 s, indicating early saturation and queuing effects from its consensus layer. InterUSS maintained p90 < 1 s and delivered ~10 lower message loss at 40 TPS, demonstrating superior scalability and stability for real-time UTM workloads.
\end{abstract}


\section{Introduction}

\begin{figure}[htbp]
      \centering
        \includegraphics[width=\linewidth]{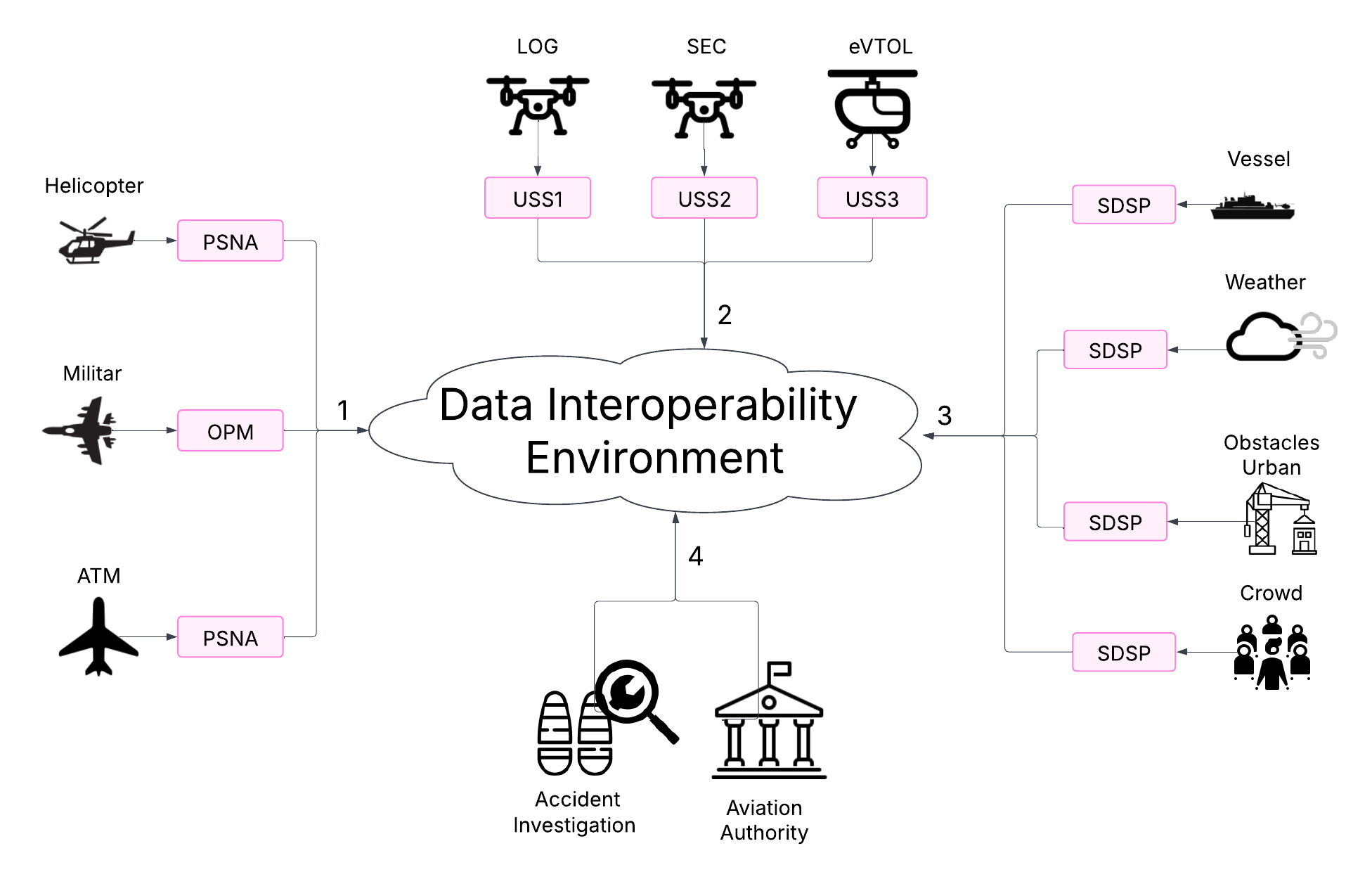}
      \caption{Data Interoperability Environment }
      \label{fig:Environment}
      \end{figure}

The rapid emergence of UAS is transforming the low-altitude airspace into a dynamic and densely occupied environment that demands new paradigms for safety\cite{FAA2023}, scalability\cite{Doole2018}, and governance\cite{u-space_conops}. To address these challenges, aviation authorities have defined distributed UTM architectures that coordinate autonomous flight operations through digital and interoperable services \cite{faa_faa_71eb3606,u-space_conops} as showed in Figure \ref{fig:Environment}. 

These concepts first proposed in the NASA UTM  \cite{nasa_tcl4} and European U-space \cite{u-space_conops} emphasize interoperability, performance-based services, and the coexistence of manned and unmanned traffic under shared situational awareness environments. Within these architectures, information exchange occurs among multiple stakeholders, including \textit{UAS Service Suppliers} (USS), \textit{Supplemental Data Service Providers} (SDSP), and \textit{Common Information Services} (CIS), through standardized interfaces supporting flight authorization, conformance monitoring, and network identification \cite{SESARjournal,SESAR}.
      
The distributed and safety-critical nature of UTM introduces strict requirements for data integrity\cite{Alkadi2022UAVInteroperability}, authentication\cite{Allouch2021UTMChain}, and traceability\cite{nasa_tcl4}. Conventional centralized data management systems present single points of failure and limited scalability, making them unsuitable for aeronautical-grade reliability. 
To ensure interoperability among multiple USS while avoiding centralized dependencies, the \textit{InterUSS Platform} was introduced as an open and federated data-exchange framework supporting standardized discovery and synchronization services across UTM domains \cite{interuss,interuss-dss-concepts}. However, it still relies on trust between participating entities and lacks native mechanisms for data immutability\cite{Allouch2021UTMChain}, provenance verification, and decentralized consensus, which are essential for safety-critical operations. To overcome these limitations, recent research has proposed the use of \textit{blockchain technology} as an enabling layer for data sharing, coordination, and event auditability across distributed stakeholders \cite{Allouch2021UTMChain,Keith2023BlockchainUTM,Baptista2024DFly}.

Permissioned ledgers based on frameworks such as \textit{Hyperledger Fabric} (Fabric) have been employed to record flight authorizations and mission data securely while ensuring compliance with privacy and governance policies\cite{freeman_oir}. Such systems introduce immutability, decentralized validation, and non-repudiation capabilities to UTM services, enhancing the trustworthiness of operational exchanges.

In Europe, studies have proposed blockchain-based U-space architectures where services like \textit{Flight Authorization} (FA), \textit{Network Identification} (NI), and \textit{Common Information} (CI) run as smart contracts among USSPs\cite{rakotonarivo_uspace}. Other works have investigated decentralized coordination mechanisms for safety and auditability in distributed UTM infrastructures \cite{Hamissi2023}, reinforcing the potential of blockchain to complement the federated data-sharing models already adopted in aviation. These contributions converge toward a vision of a distributed, regulation-compliant digital infrastructure capable of supporting high-density air mobility while maintaining accountability and operational integrity.

Building upon these insights, this work benchmarks two distributed architectures aligned with current UTM regulatory frameworks: (1) the \textit{InterUSS Platform} maintained by the \textit{Linux Foundation} \cite{interuss}, representing the state-of-practice federated approach; and (2) a \textit{Hyperledger Fabric}-based private ledger, representing a blockchain-driven alternative. By measuring transaction throughput, latency distribution, and message loss under incremental load conditions, the analysis assesses their ability to satisfy aeronautical performance constraints. 

%The comparative evaluation provides quantitative evidence of how architectural and consensus-layer differences impact scalability, responsiveness, and reliability for real-time UTM workloads---a critical step toward certifiable distributed infrastructures supporting future urban air mobility operations.

The remainder of this paper is organized as follows: Section~\ref{sec:related} presents recent research on distributed UTM systems. Section~\ref{sec:description} describes the UTM System Architecture Overview. Section~\ref{sec:benchmark} describes the methodology for performance evaluation experiments, workload  and Section~\ref{sec:results} discusses the results of the performance evaluation in terms of throughput, latency and success rate. Finally, Section~\ref{sec:conclusion} concludes the study and outlines future directions for research on scalable distributed UTM solutions.

%%%%%%%%%%%%%%%%%%%%%%%%%%%%%%%%%%%%%%%%%%%%%%%%%%%%%%%%%%%%%%%%%%%%%%%
\section{Related Works}
\label{sec:related}
%%%%%%%%%%%%%%%%%%%%%%%%%%%%%%%%%%%%%%%%%%%%%%%%%%%%%%%%%%%%%%%%%%%%%%%
Academic initiatives have explored blockchain applications in UTM, emphasizing immutability, smart contracts, certification, and auditability \cite{Allouch2021,Keith2023,Baptista2024}. Early efforts (2017–2019) mainly targeted UAV communication security through public blockchains such as Ethereum, ensuring data integrity and autonomy in UAV networks \cite{Kapitonov2017,Liang2017}.
However, these works remained limited to conceptual designs or small-scale prototypes, with no validation against aeronautical constraints or regulatory frameworks \cite{Hossain2024}.

More recent studies shifted toward permissioned blockchains—especially Hyperledger Fabric—addressing safety and reliability for low-altitude operations \cite{Allouch2021,Keith2023}. They provided decentralized registries of flight plans and validated the feasibility of secure data sharing under constrained environments \cite{Baptista2024}.
However, despite demonstrating technical viability, these solutions were validated in isolated testbeds, lacking alignment with aviation regulations (e.g., U-space, NASA UTM) and therefore remain uncertifiable in operational contexts \cite{Rakotonarivo2024, Hamissi2023}.

Other studies evaluated scalability and conflict resolution using blockchain-based coordination to deconflict multiple UAV trajectories, achieving latency between 60 ms and 2 s for up to 1000 drones \cite{Keith2023}.
However, such ad hoc environments are detached from real regulatory or operational baselines, overlooking integration, certification, and interoperability with legacy aeronautical systems.

A distinct research line explicitly targeted regulatory alignment, introducing frameworks compliant with U-space by integrating aviation authorities as blockchain validators through an Appointed-by-Authority consensus \cite{Rakotonarivo2024}. This model ensured modularity, traceability, and “secure-by-design” governance.
However, these works intentionally excluded performance and confidentiality analyses, focusing only on architectural consistency and compliance.

Conversely, public blockchain approaches, typically basead on Ethereum with zero-knowledge proofs to ensure transparency and privacy for UTM services like drone registration and flight authorization \cite{Baptista2024}.
However, public networks face intrinsic issues of scalability, latency, and governance, making them unsuitable for real-time, safety-critical operations \cite{Hossain2024}.

Moreover, although distribution is often cited as a key requirement for future UTM infrastructures, we found no prior work that directly compares blockchain-based architectures with other distributed frameworks that are compliant with sector-specific norms and standards. A particularly relevant example is the \textit{InterUSS Platform}, maintained by the Linux Foundation, which represents a federated and standardized approach for data exchange between multiple UTM service providers without relying on blockchain \cite{interuss,interuss-dss-concepts}. Designed to align with regulatory initiatives such as NASA UTM and U-space, InterUSS provides a robust operational baseline for any solution aiming at real-world adoption.

To the best of our understanding, there remains a lack of systematic comparison between blockchain-based approaches and an established baseline like InterUSS constitutes a transversal gap across the literature , as summarized in Table~\ref{tab:lit_gap}.

\begin{table}[t]
\centering
\caption{Summary of blockchain-based UTM studies and identified gaps.}
\resizebox{0.48\textwidth}{!}{
\begin{tabular}{p{2.7cm} c c c p{3cm}}
\toprule
\textbf{Work} & \textbf{Security} & \textbf{Regulatory} & \textbf{Performance} & \textbf{Gap / Remarks} \\
\midrule
\textit{UTM-Chain (2021)} \cite{Allouch2021UTMChain} & $\checkmark$ & $\checkmark$ & $\times$ & No latency or scalability benchmarks. 

\\

\textit{U-spaceChain (2024)} \cite{rakotonarivo_uspace} & $\checkmark$ & $\checkmark$ & $\times$ & Conceptual; lacks 
quantitative evaluation. 

\\

\textit{Keith et al. (2023)} \cite{Keith2023BlockchainUTM} & $\checkmark$ & $\triangle$ & $\times$ & Simulated only; no interoperability test. 

\\

\textit{Hossain et al. (2024)} \cite{Hossain2024} & $\checkmark$ & $\times$ & $\triangle$ & Models latency; not aligned with UTM specs.

\\

\textit{Baptista et al. (2024)} \cite{Baptista2024DFly} & $\checkmark$ & $\triangle$ & $\times$ & Public chain; non-deterministic timing. 

\\

\midrule
\textbf{This work} & $\checkmark$ & $\checkmark$ & $\checkmark$ & First quantitative benchmark of DLT and Federated UTM compliance. 

\\

\bottomrule
\end{tabular}
}
\vspace{1mm}
$\checkmark$ = addressed; $\triangle$ = partial; $\times$ = not addressed;
\label{tab:lit_gap}
\end{table}

Thus, there is a pressing need for integrated evaluations that address technical performance, regulatory compliance, and interoperability across distributed paradigms \cite{Rakotonarivo2024, Baptista2024,Keith2023}. 

This work seeks to address this gap by conducting a comparative analysis of regulatory alignment and operational performance between a permissioned blockchain-based UTM architecture and the federated reference implementation provided by InterUSS.

%To this end, we executed a large-scale workload comprising up to 2,000 simulated operation requests (\textit{OperationalIntentReferences}) under transaction rates ranging from 10 to 50 TPS. Each configuration was replicated across multiple trials to ensure statistical robustness, measuring latency, throughput, and loss under controlled network and computational conditions. The experiments were designed to emulate realistic UAS operation patterns derived from official Brazilian traffic data, providing a reproducible benchmark for evaluating aeronautical-grade distributed systems.

%%%%%%%%%%%%%%%%%%%%%%%%%%%%%%%%%%%%%%%%%%%%%%%%%%%%%%%%%%%%%%%%%%%%%%%
\section{UTM System Architecture Overview  }\label{sec:description}
%%%%%%%%%%%%%%%%%%%%%%%%%%%%%%%%%%%%%%%%%%%%%%%%%%%%%%%%%%%%%%%%%%%%%%%

% TODO: 2
% how is the network

% TODO: 1
% kpi found in TIM's 5G core network

The prevailing approach for managing low-altitude airspace is based on a distributed architecture, where multiple UAS Service Suppliers (USSs) operate in coordination while maintaining autonomy over their own users. This decentralized model has been promoted through regulatory frameworks such as the FAA's UTM ConOps \cite{faa_faa_71eb3606} and Europe's U-space initiative \cite{u-space_conops}, both of which emphasize the role of digital infrastructure and data interoperability for scalable airspace management.

An Airspace Reservation a represents a formal declaration of a planned operation, including key parameters such as geographical volumes, time windows, flight intent, and priority \cite{ASTM_F3548}\cite{ASTM_F3411}. The creation of an Operational Intent Reference (OIR) is a core process in the management of low-altitude airspace access within Unmanned Aircraft System Traffic Management frameworks.  

When a UAS operator intends to execute a mission, the USS must initiate a request to create an OIR by submitting this information to the DSS, which checks for potential conflicts with existing OIRs from other operators. If no conflicts are detected, the OIR is registered and made discoverable to other USSs to ensure strategic deconfliction and shared situational awareness. This process is fundamental for enabling equitable and safe access to congested low-altitude airspace, where numerous operations may overlap both spatially and temporally. As such, the ability to create, update, and delete OIRs in near real-time is critical for maintaining dynamic coordination and ensuring the integrity of autonomous or remotely piloted drone operations \cite{faa_faa_71eb3606}.

A central element of this architecture is the Discovery and Synchronization Service (DSS), which acts as a decentralized registry to enable USSs to share OIRs and awareness without storing flight data \cite{ASTM_F3411}.

The most prominent implementation of the DSS is the InterUSS Platform, an open-source solution initially proposed by Wing (an Alphabet company), Uber, and AirMap, and now governed by the Linux Foundation \cite{interuss}. The platform provides foundational services for federated coordination, including conflict detection, volume reservation, and operation notification, enabling seamless information exchange across competitive service providers \cite{nasa_tcl4}. However, while the InterUSS DSS enables standardized and efficient coordination, its trust model relies on correct behavior from participating USSs. The system lacks native immutability and verifiable audit trails, making accountability dependent on external oversight \cite{interuss,ASTM_F3411}.  

A \textit{blockchain-by-design} approach addresses these limitations by recording OIR transactions—creation, update, and deletion—as immutable, time-stamped entries validated by consensus. This ensures integrity, non-repudiation, and shared truth across the network without centralized trust \cite{Allouch2021UTMChain,Keith2023BlockchainUTM,Hossain2024,Baptista2024DFly}. Smart contracts can also automate OIR validation and conflict resolution, embedding compliance and transparency directly into the protocol \cite{rakotonarivo_uspace,freeman_oir}.  

In this view, a blockchain-enabled DSS functions as a verifiable state machine rather than a passive registry, inherently providing traceability, resilience, and trust among competing service providers \cite{Hamissi2023,Baptista2024DFly}.

\begin{figure}[h]
    \centering
    \includegraphics[width=\linewidth]{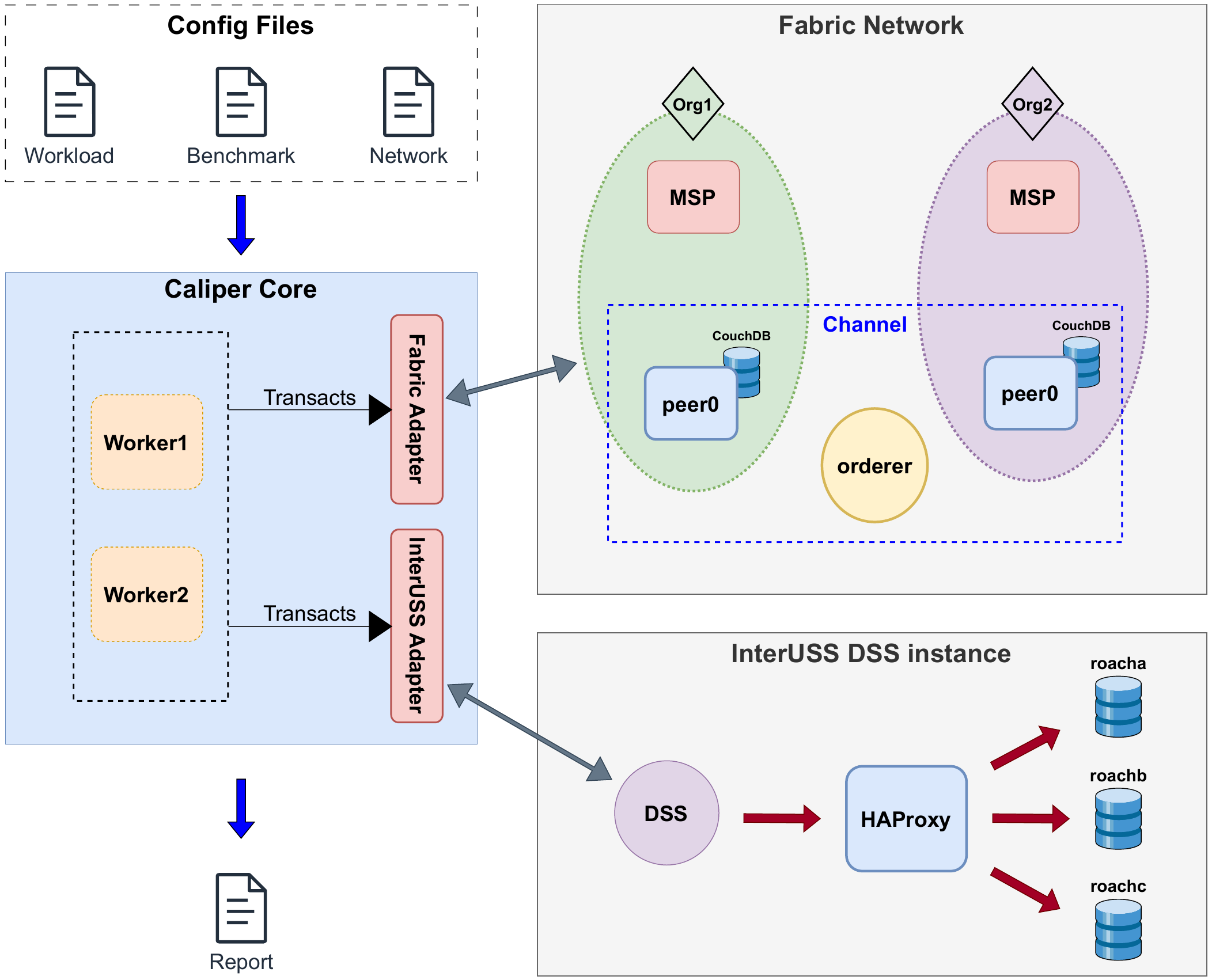}
    \caption{Architecture of the environment used in the experiments.}
    \label{fig:Experimentação}
\end{figure}

\section{Benchmark Design and Scope}
\label{sec:benchmark}

To provide a fair and reproducible performance comparison between federated and blockchain-based UTM implementations, the benchmark was designed to isolate a single and critical function common to both architectures: the \textit{Operational Intent Reference} (OIR) registration process. This operation represents the atomic unit of coordination among UAS Service Suppliers (USSs), being triggered whenever a new mission is submitted, updated, or withdrawn within a managed airspace volume \cite{ASTM_F3411,freeman_oir}.  

In the baseline implementation, the \textit{InterUSS Discovery and Synchronization Service (DSS)} was deployed following its open-source reference configuration maintained by the Linux Foundation \cite{interuss}. Each OIR submission was executed through the \textit{createOperationalIntentReference} REST endpoint, which handles request validation, conflict checking, and synchronization across participating USS nodes via the gRPC-based discovery protocol.  

For the blockchain-based scenario, we implemented an equivalent workflow using \textit{Hyperledger Fabric}. A dedicated \textit{chaincode} replicated the OIR registration logic, with identical input parameters (geographical coordinates, radius, altitude, start and end times, and priority). Each transaction was submitted to the ordering service and committed to the distributed ledger through Fabric’s endorsement and validation phases, ensuring deterministic execution and immutability.  

By maintaining parity between both implementations, this benchmark focuses exclusively on the latency, throughput, and reliability of OIR registration under incremental load conditions. The objective is to quantify how architectural differences---namely the REST-based federated synchronization of InterUSS versus the consensus-based transaction model of Hyperledger Fabric---affect aeronautical performance constraints such as timeliness and scalability.  

%The results reported in Section~\ref{sec:results} present the comparative behavior of both systems under equivalent workloads, highlighting the trade-offs between federation efficiency and ledger-level assurance mechanisms.

\subsection{Experiment Setup}

To assess the performance of both InterUSS-DSS and blockchain-based approaches for managing access to low-altitude airspace, we designed a comparative experiment centered on the creation of Operational Intent References (OIRs) through the \textit{createOperationalIntentReference} endpoint, which emulates write-intensive operations in UTM environments. Performance evaluation was conducted using \textit{Hyperledger Caliper}\footnote{https://hyperledger-caliper.github.io/caliper}, a benchmarking framework capable of measuring transaction success rate, throughput, latency, and computational resource utilization across heterogeneous distributed systems.

Figure~\ref{fig:Experimentação} illustrates the experimental architecture. In the Hyperledger Fabric setup, we deployed the official \textit{test network}\footnote{https://github.com/hyperledger/fabric-samples/tree/main/test-network}, composed of two organizations, each hosting a single \textit{peer} node, and one {orderer} service—interconnected within the same \textit{channel}. The ledger state was persisted in \textit{CouchDB}, enabling JSON-based data modeling and rich queries directly on-chain.  

For the InterUSS environment, we deployed a local setup defined in the official InterUSS DSS repository\footnote{https://github.com/interuss/dss/blob/master/build/dev/haproxy\_local\_setup.sh}, which runs on Docker and includes five containers: one for the DSS application, one for HAProxy (responsible for load balancing of the \textit{CockroachDB} cluster), and three others for the \textit{CockroachDB} nodes A, B, and C. To ensure interoperability with Caliper, we developed a dedicated \textit{InterUSS Adapter} that enables the generation of uniform workloads and the collection of comparable metrics via the InterUSS REST API.  

In both configurations, two concurrent \textit{workers} were deployed to submit transactions in parallel to each System Under Test (SUT) through their respective adapters, ensuring equivalent load conditions. After each test round, Caliper automatically generated detailed performance reports summarizing throughput, latency distribution, and system resource usage for both implementations.

\begin{figure}[h]
    \centering
    \includegraphics[width=\linewidth]{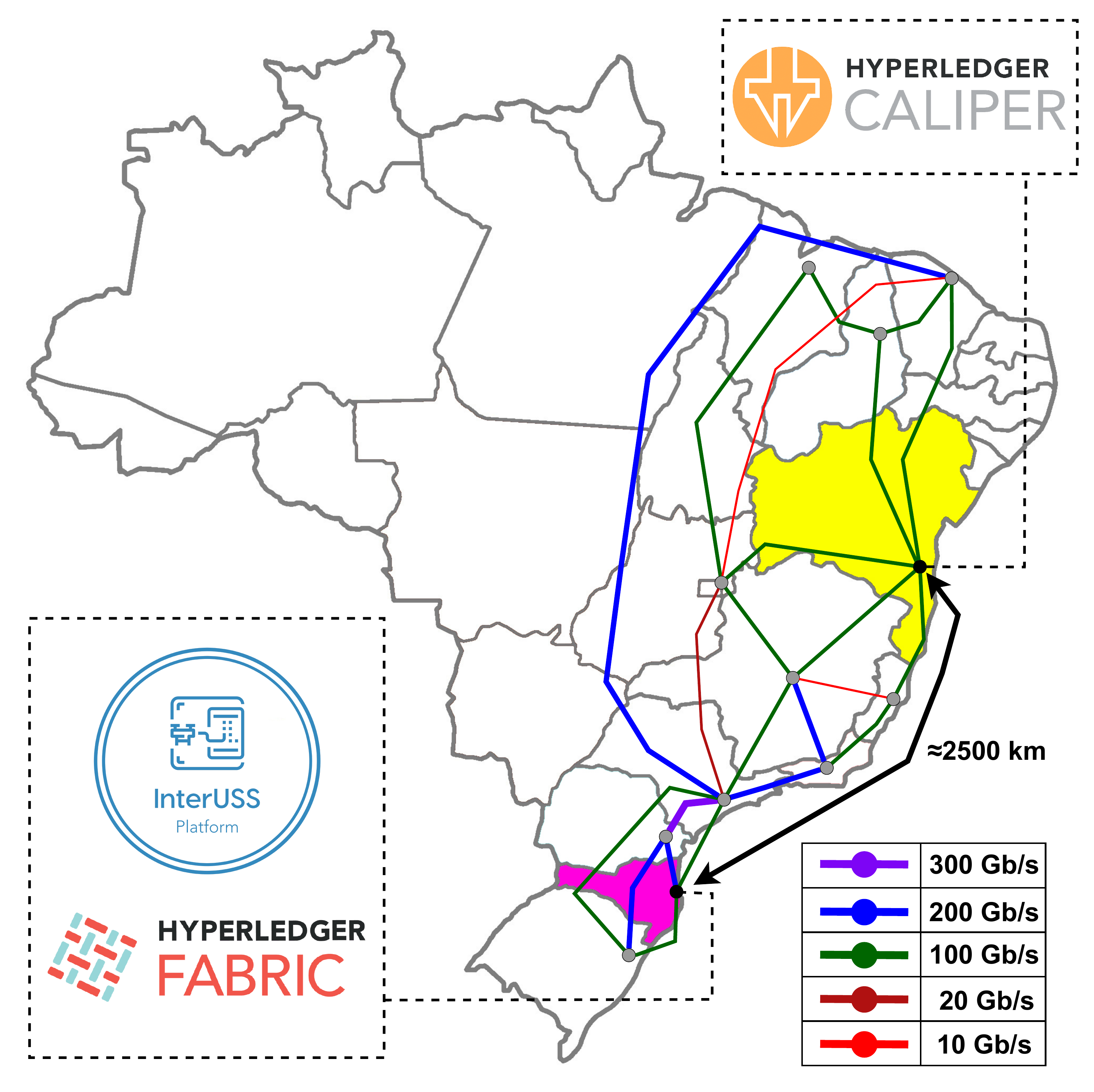}
    \caption{Partial view of the RNP’s Rede Ipê backbone infrastructure, showing links with capacities ranging from 10 to 300 Gb/s. Benchmark tests were executed remotely over this infrastructure, involving servers distributed across different Brazilian states separated by approximately 2500 km (\protect\url{https://redeipe.rnp.br}).}
    \label{fig:BackboneRNP}
\end{figure}

As shown in Figure~\ref{fig:BackboneRNP}, both Systems Under Test (SUTs) - Hyperledger Fabric and InterUSS - were hosted on the same virtual machine hosted on a server located in the state of Santa Catarina, Brazil, while the benchmarking tool Hyperledger Caliper was executed from a separate virtual machine in another server located in the state of Bahia, Brazil. Both servers featured identical hardware infrastructure composed of two Intel® Xeon® Gold 6526Y processors (2.8 GHz, 16 cores / 32 threads, 37.5 MB cache, DDR5-5200, Turbo, HT, 195 W TDP) and SAS ISE 12 Gbps 10K RPM 2.5-inch hot-plug drives. Each virtual machine was configured with 4 vCPUs, 8 GB of RAM, and 128 GB of virtual disk space. This separation between Caliper and the SUT environment was maintained to prevent the benchmarking tool from consuming computational resources of the tested systems and to ensure isolated performance measurements, and to simulate a realistic scenario in which requests could be performed remotely.

\begin{figure*}[ht]
\centering
\subfloat[Median latency (p50)]{
  \includegraphics[width=0.48\textwidth]{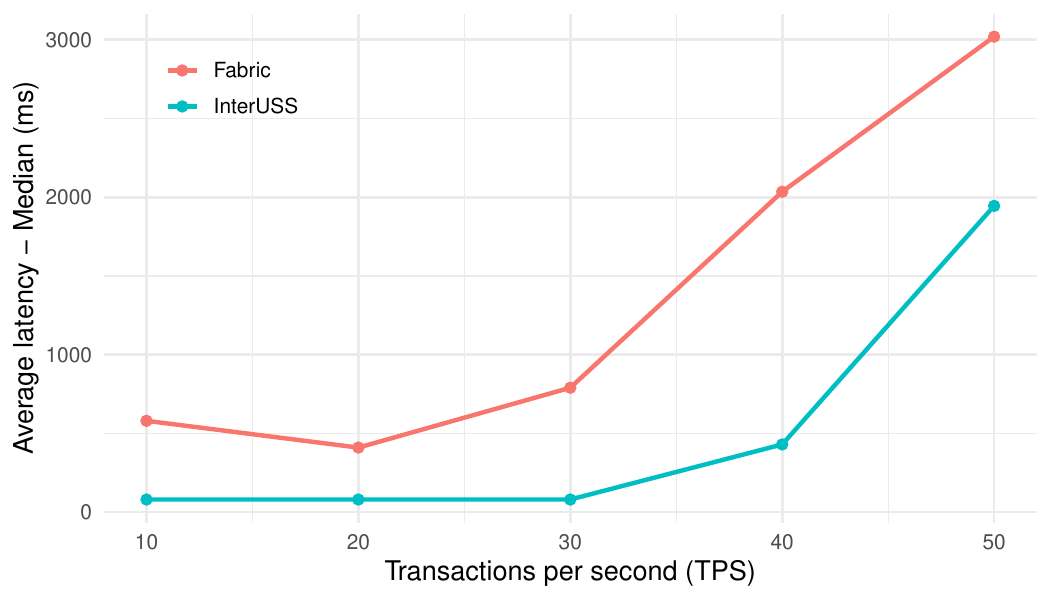}
}
\hspace{1pt}
\subfloat[P90 latency]{
  \includegraphics[width=0.48\textwidth]{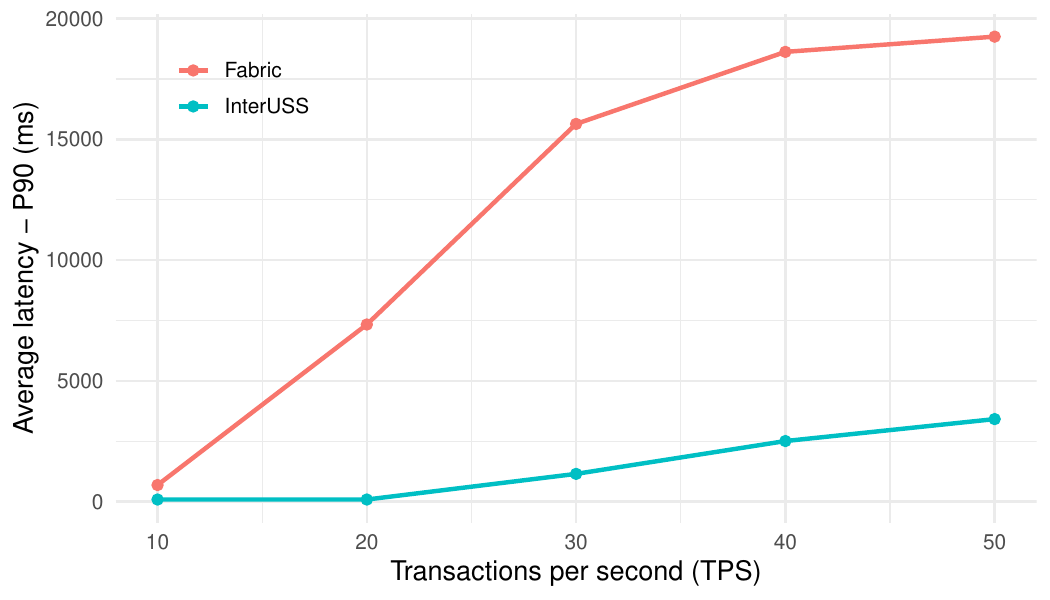}
}
\hspace{1pt}
\subfloat[Throughput]{
  \includegraphics[width=0.48\textwidth]{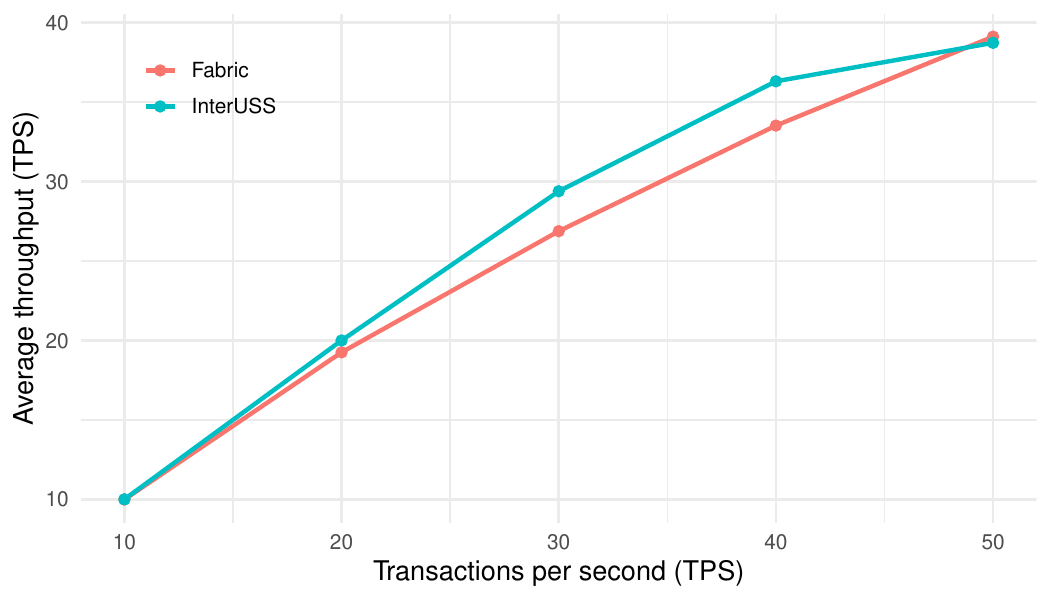}
}
\hspace{1pt}
\subfloat[Loss rate]{
  \includegraphics[width=0.48\textwidth]{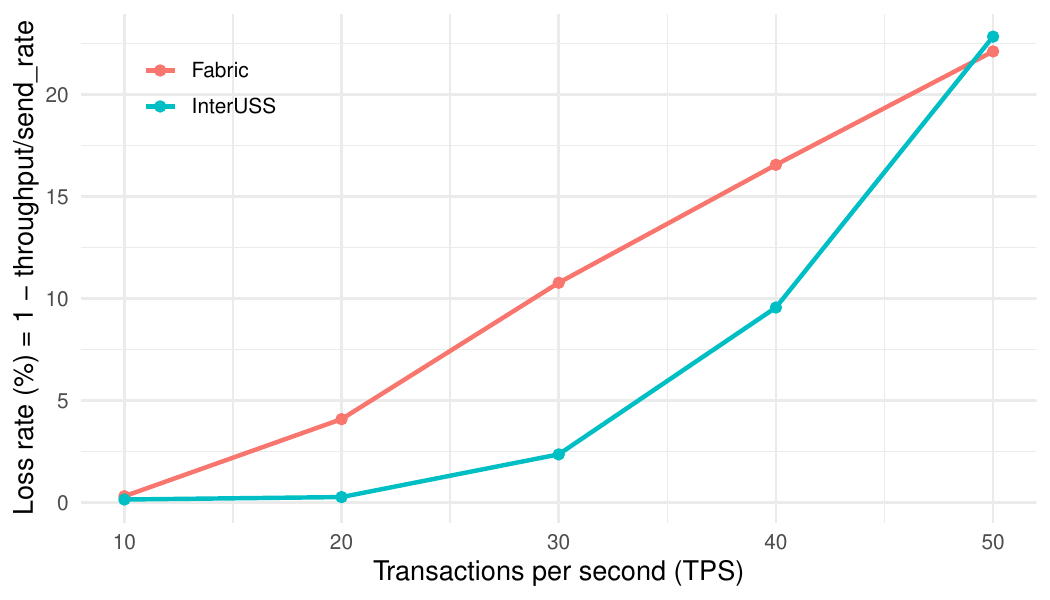}
}

\caption{
Comparative performance of Hyperledger Fabric and InterUSS platforms under increasing workload (10–50 TPS).
InterUSS maintains sub-second latency up to 30 TPS, while Fabric shows exponential degradation and higher p90 dispersion.
Both systems saturate around 50 TPS, with Fabric showing greater throughput loss.
}
\label{fig:latency_scalability}
\end{figure*}

\subsection{Workload}
To ensure the realism of our performance evaluation, we defined the experimental workload based on both theoretical modeling and real operational data. The theoretical foundation follows the approach proposed by Doole et al. (2019), who modeled urban drone delivery demand in Paris to estimate future UAM traffic density. Their analysis projected up to 100,000 operations per hour in metropolitan areas by 2035, equivalent to approximately 27.8 transactions per second, assuming each operation request represents a unique UTM transaction. To align with actual UAS operational behavior, we calibrated these estimates using 2024 data provided by the Brazilian Department of Airspace Control, which recorded 409,000 UAS operation requests throughout the year in all Brazilian territory, with a peak average demand of 33,434 requests per hour, corresponding to 9.28 transactions per second.

In this context, we evaluated different transaction volumes, ranging from 100 to 2,000 total transactions in increments of 100, and varied the transaction send rate (TPS – transactions per second) from 10 to 50, increasing by steps of 10. To prevent conflicts between OIRs and ensure a fair evaluation, we generated OIRs with intentionally staggered time windows, thereby eliminating spatio-temporal overlap between operations. Each test scenario was executed ten times to ensure statistical robustness, and the arithmetic mean of the results was used for performance comparison. 

\section{Results} \label{sec:results}

%\begin{figure*}[t]
%\centering
%\subfloat[Median latency (p50)]{
%  \includegraphics[width=0.22\textwidth]{figs/icc/latency_curve_median_vs_tps_filtered.pdf}
%}
%\hspace{2pt}
%\subfloat[P90 latency]{
%  \includegraphics[width=0.22\textwidth]{figs/icc/latency_curve_p90_vs_tps_filtered.pdf}
%}
%\hspace{2pt}
%\subfloat[Throughput]{
%  \includegraphics[width=0.22\textwidth]{figs/icc/throughput_vs_tps_filtered.pdf}
%}
%\hspace{2pt}
%\subfloat[Loss rate]{
%  \includegraphics[width=0.22\textwidth]{figs/icc/loss_vs_tps_filtered.pdf}
%}

%\caption{
%Comparative performance of Hyperledger Fabric and InterUSS platforms under increasing workload (10–50 TPS).
%InterUSS maintains sub-second latency up to 30 TPS, while Fabric shows exponential degradation and higher p90 dispersion.
%Both systems saturate around 50 TPS, with Fabric showing greater throughput loss.
%}
%\label{fig:latency_scalability}
%\end{figure*}

Figure \ref{fig:latency_scalability} presents the latency behavior of both platforms under incremental workload conditions ranging from 10 to 50 transactions per second (TPS). 

The results indicate a clear divergence in scalability behavior between the two implementations. Up to 30 TPS, the InterUSS platform maintains a nearly constant median latency around 80–100 ms, whereas the Fabric architecture exhibits an exponential growth pattern, with the median latency rising from ~580 ms at 10 TPS to ~790 ms at 30 TPS. Above 30 TPS, both systems experience degradation, but the magnitude of deterioration is considerably more pronounced for Fabric, reaching median latencies above 3 s at 50 TPS.
\begin{figure}[ht]
\centering
\includegraphics[width=\linewidth]{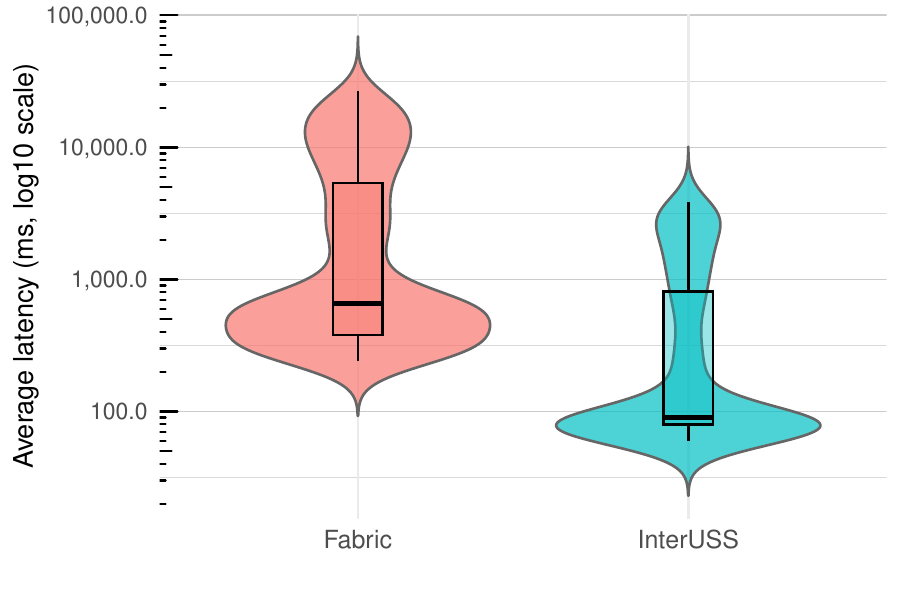}
\caption{
Distribution of average latency by platform under filtered TPS conditions (log10 scale).
Fabric shows higher latency dispersion and longer tails, while InterUSS maintains tighter and lower latency ranges.
}
\label{fig:latency_violin}
\end{figure}

The tail behavior, represented by the p90 latency metric, reinforces this trend. While InterUSS maintains a p90 below 1 s up to 30 TPS and increases moderately thereafter, Fabric shows p90 values exceeding 7 s at 20 TPS and 15 s at 30 TPS, revealing substantial instability under moderate to high transaction rates. This phenomenon suggests that the ordering and endorsement mechanisms in Fabric, which involve consensus and cryptographic validation, introduce significant queuing effects when the system approaches saturation, contrasting with the lighter REST-based, publish/subscribe interaction model of InterUSS (Fig.~\ref{fig:latency_violin}).

In terms of throughput, both architectures exhibit comparable performance for moderate loads. Up to 30 TPS, the achieved throughput closely follows the target send rate, indicating effective transaction processing and minimal queuing delay. However, at 40 TPS and above, the systems start to saturate. The throughput of Fabric drops to ~33.5 TPS (out of 40 sent), corresponding to a 16.6 \% loss rate, whereas InterUSS achieves ~36.3 TPS with 9.6 \% loss. At 50 TPS, both platforms converge to an approximate 22 \% loss rate, suggesting that this threshold marks the onset of resource saturation for both implementations.

Overall, these results highlight three major findings:

\begin{itemize}

\item Latency scalability: InterUSS presents lower and more predictable latency profiles, maintaining sub-second response times up to 30 TPS, while Fabric exhibits early exponential degradation.

\item Tail stability: The p90 latency growth in Fabric indicates limited suitability for real-time or near-real-time UTM operations under high concurrency, where deterministic timing is critical.

\item Throughput resilience: Although both architectures reach their saturation point around 50 TPS, InterUSS sustains higher throughput with lower packet loss and narrower latency dispersion.

\end{itemize}

Additional system-level metrics, including CPU, memory, disk, and network utilization, were also monitored during the benchmarking. However, these remained stable across both scenarios, indicating that the observed performance differences primarily stem from architectural and protocol-level factors rather than hardware resource limitations. 

%From a UTM perspective, such behavior implies that Fabric may be better suited for auditing, traceability, or asynchronous coordination tasks, whereas InterUSS-based architectures are more appropriate for operational data exchange and tactical decision support, where responsiveness and predictability are paramount. Future work should investigate hybrid configurations, combining DLT-based immutability for governance layers with lightweight RESTful brokers for tactical airspace interactions.

%%%%%%%%%%%%%%%%%%%%%%%%%%%%%%%%%%%%%%%%%%%%%%%%%%%%%%%%%%%%%%%%%%%%%%%
\section{Conclusion and Future Works}\label{sec:conclusion}
%%%%%%%%%%%%%%%%%%%%%%%%%%%%%%%%%%%%%%%%%%%%%%%%%%%%%%%%%%%%%%%%%%%%%%%
This work analyzed smart city drones solutions: the federated InterUSS platform and a permissioned blockchain based on Hyperledger Fabric. InterUSS sustained sub-second latency up to 30 TPS with stable tails, while Fabric exceeded 3\,s median latency and 15\,s p90 beyond this load, saturating near 50 TPS. These results show that current permissioned-ledger designs cannot meet aeronautical real-time constraints. A hybrid model that couples federated coordination with DLT-based auditability emerges as a more practical path for scalable, regulation-compliant UTM backends. In future work we plan extend the benchmark to mixed workloads and larger validator sets, and explore hybrid designs using blockchain oracles and Besu-based smart contracts in UTM.

\section*{Acknowledgements}

CAPES, FAPESP (\#2020/09850-0, \#2022/00741-0, \#2024/21006-1), CNPq, RNP/ILÍADA e Divisão de Pesquisas do Instituto de Controle do Espaço Aéreo (ICEA/FAB).
\bibliographystyle{IEEEtran}
\bibliography{IEEEabrv,references}

\begin{thebibliography}{10}
\providecommand{\url}[1]{#1}
\csname url@samestyle\endcsname
\providecommand{\newblock}{\relax}
\providecommand{\bibinfo}[2]{#2}
\providecommand{\BIBentrySTDinterwordspacing}{\spaceskip=0pt\relax}
\providecommand{\BIBentryALTinterwordstretchfactor}{4}
\providecommand{\BIBentryALTinterwordspacing}{\spaceskip=\fontdimen2\font plus
\BIBentryALTinterwordstretchfactor\fontdimen3\font minus \fontdimen4\font\relax}
\providecommand{\BIBforeignlanguage}[2]{{%
\expandafter\ifx\csname l@#1\endcsname\relax
\typeout{** WARNING: IEEEtran.bst: No hyphenation pattern has been}%
\typeout{** loaded for the language `#1'. Using the pattern for}%
\typeout{** the default language instead.}%
\else
\language=\csname l@#1\endcsname
\fi
#2}}
\providecommand{\BIBdecl}{\relax}
\BIBdecl

\bibitem{FAA2023}
{Federal Aviation Administration}, ``Utm field test (uft) final report, version 1.0,'' Federal Aviation Administration, Tech. Rep., Nov. 2023.

\bibitem{Doole2018}
M.~Doole, J.~Ellerbroek, and J.~Hoekstra, ``\BIBforeignlanguage{English}{Drone delivery: Urban airspace traffic density estimation},'' in \emph{\BIBforeignlanguage{English}{8th SESAR Innovation Days, 2018}}, 2018.

\bibitem{u-space_conops}
\BIBentryALTinterwordspacing
EUROCONTROL, ``U-space conops and architecture (edition 4),'' Tech. Rep., 2023. [Online]. Available: \url{https://www.sesarju.eu/sites/default/files/documents/reports/U-space%20CONOPS%204th%20edition.pdf}
\BIBentrySTDinterwordspacing

\bibitem{faa_faa_71eb3606}
\BIBentryALTinterwordspacing
{Federal Aviation Administration}, ``{Unmanned Aircraft System (UAS) Traffic Management (UTM) Concept of Operations V1.0},'' Tech. Rep., 2018. [Online]. Available: \url{https://www.faa.gov/sites/faa.gov/files/2022-08/UTM_ConOps_v2.pdf}
\BIBentrySTDinterwordspacing

\bibitem{nasa_tcl4}
\BIBentryALTinterwordspacing
{NASA UTM Project Team}, ``Nasa utm technical capability level 4 (tcl4) flight demonstrations final report,'' NASA, Tech. Rep. NASA/TP–20205010302, 2020. [Online]. Available: \url{https://ntrs.nasa.gov/citations/20205010302}
\BIBentrySTDinterwordspacing

\bibitem{SESARjournal}
T.~Boli\'c and P.~Ravenhill, ``Sesar: The past, present, and future of european air traffic management research,'' \emph{Engineering}, vol.~7, pp. 448--451, 4 2021.

\bibitem{SESAR}
{SESAR Joint Undertaking}, ``U-space concept of operations,'' \url{https://www.sesarju.eu/U-space}, 2019, accessed: 2025-04-30.

\bibitem{Alkadi2022UAVInteroperability}
R.~Alkadi, N.~Alnuaimi, C.~Y. Yeun, and A.~Shoufan, ``Blockchain interoperability in unmanned aerial vehicles networks: State-of-the-art and open issues,'' \emph{IEEE Access}, vol.~10, pp. 14\,463--14\,479, 2022.

\bibitem{Allouch2021UTMChain}
A.~Allouch, O.~Cheikhrouhou, A.~Koub{\^a}a, K.~Toumi, M.~Khalgui, and T.~Nguyen~Gia, ``{UTM-Chain}: Blockchain-based secure unmanned traffic management for internet of drones,'' \emph{Sensors}, vol.~21, no.~9, p. 3049, 2021.

\bibitem{interuss}
{Linux Foundation}, ``Interuss platform project,'' \url{https://www.interuss.org}, 2023, accessed: 2025-04-30.

\bibitem{interuss-dss-concepts}
{InterUSS Platform}, ``Discovery and synchronization service concepts,'' \url{https://github.com/interuss/dss/blob/master/concepts.md}, 2022, accessed: 2025-04-30.

\bibitem{Keith2023BlockchainUTM}
A.~Keith, T.~Sangarapillai, A.~Almehmadi, and K.~El-Khatib, ``A blockchain-powered traffic management system for unmanned aerial vehicles,'' \emph{Applied Sciences}, vol.~13, no.~19, p. 10950, 2023.

\bibitem{Baptista2024DFly}
F.~Baptista, M.~Dehez-Clementi, and J.~Detchart, ``{DFly}: A publicly auditable and privacy-preserving uas traffic management system on blockchain,'' \emph{Drones}, vol.~8, no.~8, p. 410, 2024.

\bibitem{freeman_oir}
\BIBentryALTinterwordspacing
K.~Freeman, N.~Gillem, A.~Jones, and N.~Sharma, \emph{A Blockchain Case Study for Urban Air Mobility Operational Intent}, 2023. [Online]. Available: \url{https://arc.aiaa.org/doi/abs/10.2514/6.2023-3401}
\BIBentrySTDinterwordspacing

\bibitem{rakotonarivo_uspace}
B.~Rakotonarivo and M.~Bronz, ``U-spacechain: A decentralized approach to unmanned traffic management services provision,'' in \emph{2024 IEEE International Conference on Blockchain (Blockchain)}, 2024, pp. 450--457.

\bibitem{Hamissi2023}
A.~Hamissi, A.~Dhraief, and L.~Sliman, ``On safety of decentralized unmanned aircraft system traffic management using blockchain,'' in \emph{Proc. of the 32nd IEEE International Conference on Enabling Technologies: Infrastructure for Collaborative Enterprises (WETICE)}, 2023, pp. 1--6.

\bibitem{Allouch2021}
A.~Allouch, O.~Cheikhrouhou, A.~Koub\^{a}a, K.~Toumi, M.~Khalgui, and T.~Nguyen~Gia, ``{UTM-Chain}: Blockchain-based secure unmanned traffic management for internet of drones,'' \emph{Sensors}, vol.~21, no.~9, p. 3049, 2021.

\bibitem{Keith2023}
A.~Keith, T.~Sangarapillai, A.~Almehmadi, and K.~El-Khatib, ``A blockchain-powered traffic management system for unmanned aerial vehicles,'' \emph{Applied Sciences}, vol.~13, no.~19, p. 10950, 2023.

\bibitem{Baptista2024}
F.~Baptista, M.~Dehez-Clementi, and J.~Detchart, ``{DFly}: A publicly auditable and privacy-preserving uas traffic management system on blockchain,'' \emph{Drones}, vol.~8, no.~8, p. 410, 2024.

\bibitem{Kapitonov2017}
A.~Kapitonov, S.~Lonshakov, A.~Krupenkin, and I.~Berman, ``Blockchain-based protocol of autonomous business activity for multi-agent systems consisting of uavs,'' in \emph{Proc. 2017 Workshop on Research, Education and Development of Unmanned Aerial Systems (RED-UAS)}, 2017, pp. 84--89.

\bibitem{Liang2017}
X.~Liang, J.~Zhao, S.~Shetty, and D.~Li, ``Towards data assurance and resilience in iot using blockchain,'' in \emph{Proc. IEEE Military Communications Conference (MILCOM)}, 2017, pp. 261--266.

\bibitem{Hossain2024}
M.~I. Hossain, M.~Tahtali, U.~Turhan, and K.~Biswas, ``Blockchain integration in uav networks: Performance metrics and analysis,'' \emph{Sensors}, vol.~24, no.~23, p. 7813, 2024.

\bibitem{Rakotonarivo2024}
B.~Rakotonarivo and M.~Bronz, ``U-spacechain: A decentralized approach to unmanned traffic management services provision,'' in \emph{Proc. IEEE Int. Conf. on Blockchain (Blockchain)}, 2024, pp. 450--456.

\bibitem{ASTM_F3548}
\BIBentryALTinterwordspacing
ASTM, ``Astm f3548-21: Standard specification for uas traffic management (utm) uas service supplier (uss) interoperability,'' West Conshohocken, PA, USA, Tech. Rep., 2021. [Online]. Available: \url{https://www.astm.org/f3548-21.html}
\BIBentrySTDinterwordspacing

\bibitem{ASTM_F3411}
\BIBentryALTinterwordspacing
------, ``Astm f3411-22: Standard specification for remote id and tracking,'' West Conshohocken, PA, USA, Tech. Rep., 2022. [Online]. Available: \url{https://www.astm.org/f3411-22.html}
\BIBentrySTDinterwordspacing

\end{thebibliography}

\end{document}